\documentclass[prb,aps,twocolumn,showpacs,superscriptaddress,floatfix]{revtex4-1}

\usepackage{amsmath}
\usepackage{amssymb}
\usepackage{amsfonts}
\usepackage{mathptmx}

\usepackage{graphicx}
\usepackage{epstopdf}
\usepackage{dcolumn}
\usepackage{bm}

\DeclareMathOperator{\re}{\mathop{\mathrm{Re}}}
\DeclareMathOperator{\im}{\mathop{\mathrm{Im}}}
\DeclareMathOperator{\arctanh}{arctanh}

\begin{document}

\title{ Andreev current enhancement and subgap conductance of superconducting hybrid
structures in the presence of a small spin-splitting field}
\author{A.~Ozaeta}
\affiliation{Centro de F\'{i}sica de Materiales (CFM-MPC), Centro
Mixto CSIC-UPV/EHU, Manuel de Lardizabal 4, E-20018 San
Sebasti\'{a}n, Spain}
\author{A.~S.~Vasenko}
\affiliation{Institut Laue-Langevin, 6 rue Jules Horowitz, BP 156,
38042 Grenoble, France}
\author{F.~W.~J.~Hekking}
\affiliation{LPMMC, Universit\'{e} Joseph Fourier and CNRS, 25
Avenue des Martyrs, BP 166, 38042 Grenoble, France}
\author{F.~S.~Bergeret}
\affiliation{Centro de F\'{i}sica de Materiales (CFM-MPC), Centro
Mixto CSIC-UPV/EHU, Manuel de Lardizabal 4, E-20018 San
Sebasti\'{a}n, Spain}

\affiliation{Donostia International Physics Center (DIPC), Manuel
de Lardizabal 5, E-20018 San Sebasti\'{a}n, Spain}

\date{\today}

\begin{abstract}
We investigate  the subgap transport properties of  a S-F-N$_{\rm e}$ structure. Here S
(N$_{\rm e}$) is a superconducting (normal) electrode, and F is either a ferromagnet or a
normal wire in the presence of an exchange or a  spin-splitting Zeeman field
respectively.  By solving the quasiclassical equations we first analyze the behavior   of
the subgap current, known as the Andreev current,   as a function of the field strength
for different values of the voltage, temperature and length of the junction. We show
that there is a critical value of the bias voltage $V^*$ above which  the Andreev current
is enhanced by the spin-splitting  field. This unexpected behavior can be explained as
the competition between two-particle tunneling processes and decoherence mechanisms
originated from the temperature, voltage and exchange field respectively.   We also show
that at finite temperature  the Andreev current  has a peak for values of the exchange
field  close to the superconducting gap. Finally, we compute the differential conductance
and show that its measurement  can be used as an accurate way of determining the strength
of spin-splitting fields smaller than the superconducting gap.
\end{abstract}


\maketitle

{\it Introduction-} Transport properties of hybrid structures consisting of
superconducting and non-superconducting materials have been studied extensively in the
last decades \cite{Nazarov_book}. 
{ In particular, there is  a  renewed interest in the study of the subgap conductance of superconducting hybrid structures in the presence of  Zeeman-like fields in view of the presumable detection of Majorana Fermions  \cite{Kouwenhoven2012}. }
Intuitively, due to the gap $\Delta$ in the density of
states of  a superconductor the charge transport through a superconductor-normal (S-N)
metal junction    is expected to vanish for voltages smaller than  $\Delta$. However,
this is not always the case.   Experiments on  S/N  structures have shown   a finite
subgap conductance \cite{Kastalsky1991}.  This behavior was discussed theoretically  in
Refs.~\onlinecite{Hekking,VZK}.  It was shown that the conductance of a  S-N-N$_{\rm e}$
structure, where N$_{\rm e}$ denotes a normal metal electrode,   shows a peak at  a voltage
smaller than the superconducting gap\cite{footnote0} in the case of finite S/N barrier
resistances  or if  N  is a diffusive metal \cite{Hekking,VZK}. A similar behavior was
predicted if one substitutes the normal  by a ferromagnetic metal (F)
\cite{Leadbetaer1999,Seviour1999,Yokoyama2005}. In all these examples, the key mechanism
to explain the finite subgap conductance  is the Andreev reflection \cite{Andreev,
Pannetier2000} which takes place at the S/N and S/F interfaces and  allows the flow of an
electric current even for voltages smaller than the superconducting gap $\Delta$. By this
process an electron from the normal region is reflected as a hole forming a coherent
electron-hole pair which penetrates into a diffusive  normal region over distances of the order of
the thermal length $\sqrt{\mathcal{D}/T}$, where
$\mathcal{D}$ is the diffusion coefficient and $T$ is the temperature (here and below we
set $\hbar=k_B=1$). This mechanism leads to a finite condensate density in the normal
metal, i.e. to the so called superconducting proximity effect.

At a S/F interface the mechanism of charge transport is  however modified since the
incoming electron and reflected hole belong to different spin bands \cite{deJong}. Thus,
one expects  a suppression of the Andreev current by increasing the exchange field $h$ of
the ferromagnet, which is a measure for the spin-splitting at the Fermi level. In the
ferromagnet the coherence length  of the electron-hole pairs is given by the minimum
between the thermal  and the magnetic ($\sim\sqrt{\mathcal{D}/h}$) lengths. One expects
that by increasing the strength of the field $h$ the electron-hole coherence would be
suppressed and hence the subgap current reduced. As we show below,  this intuitive
picture does not hold always.

In this Letter we analyze  the Andreev current and conductance through a  S-F-N$_{\rm e}$
hybrid structure as a function of the field $h$.  Here $h$ denotes either the intrinsic
exchange field of a ferromagnet or  a spin-splitting field in a normal metal caused by
either an external magnetic field or the proximity of an insulating ferromagnet
\cite{Cottet2011}.  We focus our study on weak fields, $h\leqslant\Delta$ and $h \gtrsim
\Delta$.  We find an interesting interplay between phase-coherent diffusive propagation
of Andreev pairs due to the proximity effect and decoherence mechanisms originated from
the temperature, voltage and exchange field respectively. This interplay leads to a  non-monotonic
behavior of the transport properties as a function of $h$. For very low temperatures and
voltages  $eV\ll\Delta$  the Andreev current  decays monotonically  by increasing $h$ as
expected.  If one keeps the voltage low but now increases the temperature,  the Andreev
current shows a peak at $h\approx\Delta$.  An unexpected behavior is obtained when the
voltage exceeds some critical value $V^*$. In this case, the Andreev current is enhanced
by the field  $h$  reaching  a maximum  at $h\approx eV$.  We show that the value of
$V^*$ depends on the length of the F wire and the temperature. In particular, for
zero-temperature  and  in the  long-junction limit, i.e. when the length of F is much
larger than the coherence length,  we show that  $eV^*\approx0.56\Delta_0$, where
$\Delta_0$ is the value of $\Delta$ at $T=0$.   We also compute the subgap conductance of
the system at low temperatures and small fields $h<\Delta$.  We show that it has a peak
at  $eV = h$. Thus, transport measurements of this type can be used to determine the
strength of a weak exchange or Zeeman-like field in the nanostructure.

{\it Model and basic equations-} We consider a  ferromagnetic  wire F. Its  length,  $L$ is  smaller than the inelastic relaxation length. The wire is  attached at $x=0$ to a superconducting (S)  and at $x=L$ to a normal  (N$_{\rm e}$) electrode. As noticed above,  F can also describe a normal wire in a spin-splitting field $B$ (in which case $h = \mu_B B$, where $\mu_B$ is the Bohr magneton) or in proximity with an insulating ferromagnet \cite{Cottet2011}.  We consider the diffusive limit, i.e. we assume that the elastic scattering length is much smaller than the decay length of the superconducting condensate into the F region. 
In order to describe the transport properties of the system we compute the quasiclassical Green functions \cite{Larkin1986,Belzig}. They obey the Usadel equation \cite{Usadel1970} that in the so called $\theta-$ parametrization reads \cite{Belzig}
\begin{equation}
\partial_{xx}^2\theta_\pm=2i\frac{E\pm h}{\mathcal{D}}\sinh\theta_\pm.\label{Usadel}
\end{equation}
Here the upper (lower) index denotes the spin-up (down) component.  The normal and anomalous Green functions are given by $g_\pm=\cosh\theta_\pm$ and $f_\pm=\sinh\theta_\pm$ respectively.
Because of the high transparency of the  F/N$_{\rm e}$ interface  the functions $\theta_\pm$ vanish at $x=L$, {\it i.e.} superconducting correlations are negligible at the F/N$_{\rm e}$ interface.  {We  consider a tunneling barrier at  the S/F interface and assume that its tunneling resistance $R_T$ is  much larger than the normal resistance $R_F$ of the F layer.   Thus, by voltage-biasing  the N$_{\rm e}$  the voltage drop takes place at the  S/F tunnel interface. To leading order  in  $R_F/R_T\ll 1$ the Green functions obey  the Kupriyanov-Lukichev  boundary condition at $x=0$ \cite{KL}}
\begin{equation}
\partial_x\theta_\pm |_{x=0}=\frac{R_F}{LR_T}\sinh[\theta_\pm |_{x=0} - \theta_S],\label{BC}
\end{equation}
where $\theta_S= \arctanh (\Delta/E)$ is the superconducting bulk value of the function $\theta$.  Once the functions $\theta_\pm$ are obtained one can compute the current through the junction. In particular, we are interested in the Andreev current, i.e. the current for voltages smaller than the superconducting gap due to Andreev processes at the S/F interface. Such current  is given by the  expression \cite{VZK, VBCH}
\begin{equation}\label{I_A}
I_A=\sum_{j=\pm}\int_0^{\Delta}
\frac{n_-(E)\; dE / 2eR_T}{2W\alpha_j(E) - \sqrt{1-(E/\Delta)^2}\im^{-1}(\sinh\theta_j |_{x=0})},
\end{equation}
where $n_-(E)=\frac{1}{2}(\tanh[(E+eV)/2T]-\tanh[(E-eV)/2T])$ is the quasiparticle distribution function in the N$_{\rm e}$ electrode,  $\alpha_\pm(E)=(1/\xi)\int_0^L dx \;\cosh^{-2}[\re\theta_\pm(x)]$,  $W=\xi R_F / 2 L R_T$ is the diffusive tunneling parameter \cite{KL, Chalmers}, and $\xi = \sqrt{\mathcal{D}/2\Delta}$ is the superconducting coherence length. Eq.~(\ref{I_A}) is the expression used throughout this article in order to determine the subgap charge transport \cite{footnote1}.

\begin{figure}[t]\begin{center}
\includegraphics[scale=0.23]{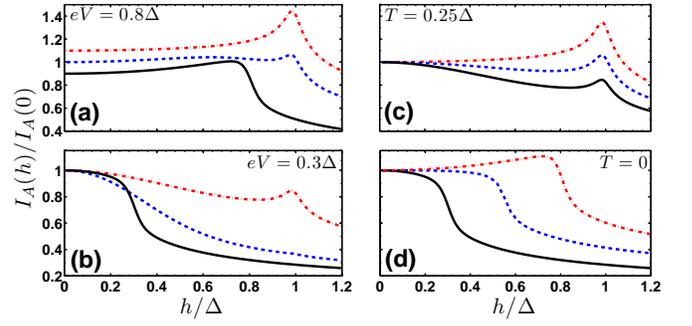}
\caption{(Color online) The $h$-dependence of the ratio $I_A(h)/I_A(0)$ for $L = 10 \xi$
and  $W = 0.007$. Left panels correspond to (a) $eV=0.8\Delta$ and (b) $eV=0.3\Delta$.
The different curves are for $T=0 $ (black solid line), $T=0.12 \Delta_0$ (blue dashed
line) and $T=0.25 \Delta_0$ (red point-dashed line).  The  right panels corresponds to (c)
$T=0.25\Delta_0$ and (d) $T=0$, while the different curves to  $eV=0.3 \Delta$ (black solid
line), $eV=0.55 \Delta$ (blue dashed line) and $eV=0.8 \Delta$ (red point-dashed line).
In  the (a) panel curves are vertically shifted with respect to each other for clarity.}
\label{a1} \vspace{-4mm}
\end{center}
\end{figure}

{\it Results-} We first compute the Andreev current  numerically by solving
Eqs.~(\ref{Usadel}-\ref{I_A}). In Fig.~\ref{a1} we show the dependence of the Andreev
current on the exchange field $h$  for different values of the bias voltage and
temperature for a ferromagnetic F wire of the length $L=10\xi$.

We consider first the zero-temperature limit. For small enough voltages ({\it e.g} $eV=0.3\Delta$, black solid line in 
 Fig. \ref{a1}b) the Andreev current decays
monotonously  with increasing $h$. This behavior  is the one expected, since  by
increasing $h$ the coherence length  of the Andreev pairs in the normal region is
suppressed, leading to a reduction of the subgap current. For  large enough voltages
({\it e.g} $eV=0.8\Delta$ in Fig.~\ref{a1}) and keeping the temperature low, the Andreev current
first increases by increasing $h$, reaches a maximum at $h\approx eV$, and then drops by
further increase of the exchange field,  as it is shown for example  by the black solid
line in Fig.~\ref{a1}a.  A common feature of all the low- temperature curves in
Fig.~\ref{a1} is the  sharp suppression of the Andreev current at $h\approx eV$.

For large enough temperatures ($T=0.25\Delta_0$ in Fig.\ref{a1}c)  one observes a peak at
$h\approx\Delta$  [Fig.~\ref{a1}(c)].  The  relative height of this peak  increases with
temperature and voltage as one sees in Figs.~\ref{a1}a and \ref{a1}c respectively. In
the case of  large enough values of  $V$ and $T$, one is able to observe both  the
enhancement of the Andreev current by increasing $h$ and the peak at $h\approx\Delta$
(see for example blue dashed line in Fig~\ref{a1}a). For values of the exchange field larger than $\Delta$,  the Andreev current decreases by increasing $h$ in all cases . In principle all  the  behaviors of
the Andreev current  can be  observed by measuring the full electric current through the
junction as the single particle current is almost independent of $h$.

In order to give a physical interpretation of these results, we first recall the details
of the process of two-electron tunneling that gives rise to subgap current~\cite{Hekking}
in diffusive systems in the absence of an exchange field. The value of this current is
governed by two competing effects. On the one hand, the origin of the subgap current is
the tunneling from the normal metal to the superconductor of two electrons with energies
$\xi_{k_1}$ and $\xi_{k_2}$, respectively and momenta $k_1$ and $k_2$, that form a Cooper pair.
This process is of the second order in tunneling and therefore involves a virtual state
with an excitation on both sides of the tunnel barrier. The relevant virtual state
energies are given by the difference $E_k - \xi_{k_1,k_2}$, where $E_k = \sqrt{\Delta^2 +
\xi_k^2}$ is the excitation energy of a quasiparticle with the momentum $k$ in the
superconductor. Typical values of $\xi_{k}$ are  $T$ or $eV$. Hence under subgap
conditions $T,eV \ll \Delta$, the virtual state energy is typically given by the
superconducting gap $\Delta$. However, when these characteristic energies become larger
and approach the value of the gap, the difference $E_k-\xi_{k_1,k_2}$ eventually vanishes. As a
result, the amplitude for two-electron tunneling increases drastically, leading to a
strong increase of the Andreev current, accompanied by the onset of single-particle
tunneling at energies above the gap $\Delta$. On the other hand, two-electron tunneling
is a coherent process: the main contribution to two-electron tunneling stems from two
nearly time-reversed electrons $k_1 \simeq -k_2$ located in an energy window of width
$\delta \epsilon \sim eV,T$ close to the Fermi energy, diffusing phase-coherently over a
typical distance $L_{coh} = \sqrt{\mathcal{D}/(\delta \epsilon)}$ in the normal metal
before tunneling~\cite{Hekking}. This coherence length decreases upon increasing the
characteristic energies $\delta \epsilon \simeq eV,T$, thereby decreasing the Andreev
current.

\begin{figure}[t]
\includegraphics[width=\columnwidth]{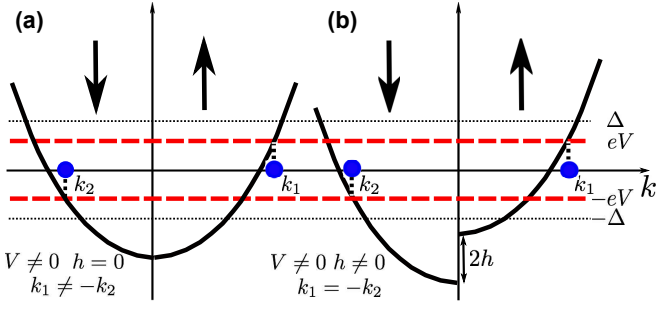}

\caption{(Color online)  Schematic energy diagram  for a non magnetic (a) and magnetic (b) metal.  The thick solid parabolas are the dispersions of free electrons with spin up ($\uparrow$) and spin-down ($\downarrow$). 
The $k$ axis corresponds to the  Fermi level in the superconductor.   We consider quasi-electrons and -holes with energies $\pm eV<\Delta$ and momentum $k_{1,2}$ . Time-reversal pairing requires that $k_1=k_2$. In case of a normal metal [panel  (a)] this condition is satisfied only for $eV=0$ while  for a ferromagnet ($h\neq0$)  if $h=eV$.}\label{par}
\end{figure}

We now turn to the effect of the exchange field $h$ on two-electron tunneling. If $h$ is
nonzero, the majority and minority spin electrons at the Fermi level are characterized by different 
wave vectors $k_{F, \pm} = k_F \mp \delta k$, where $\delta k \sim h/v_F$ and $v_F$ is the Fermi velocity.  
In Fig. \ref{par} we show a schematic energy diagram. The wave vectors
$k_{F, \pm}$ are determined by intersection between the parabolas and the $k$-axis. For a given value of $eV\lesssim\Delta$ and in the absence of an exchange field  the relevant excitations with energies $\sim \pm eV$ and wave vectors $k_{1,2}$ are not time-reversed (see Fig. \ref{par}a) and  therefore their contribution to the current  is not coherent. However upon increasing $h$, $|k_1|\rightarrow |k_2|$, {\it i.e} the relevant excitations  become more and more coherent, leading to an additional increase of two-electron tunneling. In particular when $h=eV$, $k_1=-k_2$ ({\it cf.} Fig. \ref{par}b).
If $T\rightarrow 0$  there are no occupied states for $\xi_k>eV$. 
Consequently  as soon as $h> eV$, the energy window around the Fermi level does not contain
time-reversed electrons. This leads  to the drop of the Andreev current shown for example in  Fig. \ref{a1}d.  In contrast,  for finite values of $T$,  there are thermally induced quasiparticles with energy $\sim \Delta$,  that become exactly time-reversed whenever $h=\Delta$. This leads to the maximum of the current at $h=\Delta$ when the temperature is finite ({\it cf.} Fig. \ref{a1}c). The effects are most clearly seen when plotting the
ratio $I_A(h)/I_A(0)$, as the Andreev pair decoherence effects due to temperature or
voltage are then divided out.

\begin{figure}[t]
\includegraphics[height=4.1cm,width=\columnwidth]{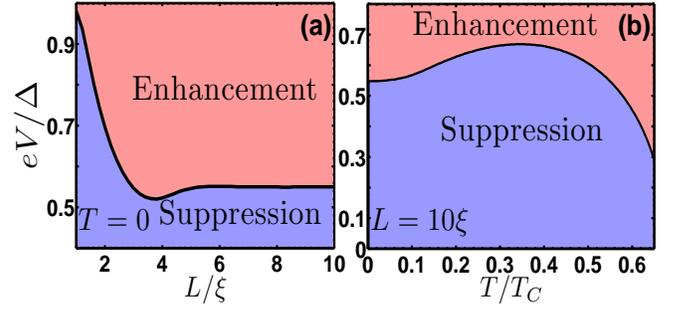}

\caption{(Color online)  Voltage-junction length (a) and voltage-temperature (b)
diagrams.  The black solid line represents the values of $eV^*/\Delta$. For the range of
parameters situated  below this line the  Andreev current decreases in the presence of a
small  exchange field (suppression), while   in the region above the line  the current
increases (enhancement).  We set $W = 0.007$ in both panels, $T=0$ in panel  (a)  and
$L=10\xi$ in panel (b).} \label{a4} \vspace{-4mm}
\end{figure}

A more quantitative understanding of the effects discussed above can be get by analyzing
some limiting cases in which simple analytical expressions for the current can be
derived.    We first focus our analysis on the zero-temperature limit.   Due to the
tunneling  barrier at the S/F interface  the proximity effect is weak and hence  one
can linearize  Eqs.~(\ref{Usadel}-\ref{BC}) with respect to $R_F/R_T \ll 1$.  After a
straightforward calculation one obtains  the Andreev current in this limit,
\begin{align}
I_A = &\frac{W \Delta_0^2}{2 e R_T}\sum_{j=\pm}\int_0^{eV} \frac{dE}{\Delta_0^2 - E^2} \nonumber
\\
&\times \re\left[ \sqrt{\frac{i \Delta_0}{E +j h}} \tanh \left( \sqrt{\frac{E + jh}{i\Delta_0}}\frac{L}{\xi} \right) \right].\label{lin_IA}
\end{align}
For  a large exchange field, $h\gg\Delta_0>eV$  one can evaluate  this expression obtaining
\begin{equation}
I_A\approx \frac{R_F \Delta_0}{8eLR_T^2}\sqrt{\frac{\mathcal{D}}{h}}\log\left[\frac{\Delta_0+eV}{\Delta_0-eV}\right].
\end{equation}
Thus, the Andreev current  decays  as $h^{-1/2}$ for  large values of $h$ in accordance
with our numerical results (see Fig.~\ref{a1}).

In the case of small values of $h$, $h\lesssim eV<\Delta_0$,  one can evaluate
Eq.~(\ref{lin_IA}) in   the  long-junction limit, i.e. when $L\gg\sqrt{D/h}$.   In this
case  the  Andreev current reads
\begin{equation}
I_A=\frac{\Delta_0 \xi R_F} {eLR_T^2}\sum_{j=\pm} \frac{\arctanh\left(\sqrt{\frac{eV+jh}{\Delta_0+jh}}\right)
+\arctan\left(\sqrt{\frac{eV-jh}{\Delta_0+jh}}\right)}{\sqrt{\Delta_0+jh}}.\label{ia_lj}
\end{equation}
This expression describes the two different behaviors obtained  in Fig.~\ref{a1} for
$h\leq eV$. For small voltages $I_A$ decreases by increasing the field $h$. However, for
large enough values of the voltage $I_A$ is enhanced by the presence of the field. From
Eq.~(\ref{ia_lj})  we can determine  the voltage $V^*$, at which the crossover between
these two behaviors takes place,  by  expanding the expression for the current up to
second order in  $h/eV\ll 1$, {\it i.e.} up to the first non-vanishing correction to
$I_A$ due to the  exchange field. This expansion leads to   the following transcendental
expression which determine the voltage $V^*$ at which the crossover takes place,
\begin{equation}
\left(\frac{\Delta_0}{eV^*}\right) ^{3/2}=\frac{3}{2}\left( \arctanh\sqrt{eV^*/\Delta_0}+\arctan\sqrt{eV^*/\Delta_0}\right).
\end{equation}
From here we get  $eV^*\approx0.56 \Delta_0$.  For  $V<V^*$ the Andreev current decays
monotonically with $h$ while for  $V>V^*$  it increases up to a maximum value at
$h\lesssim eV$. This  is in agreement with our numerical results in Fig.~\ref{a1}.

For an  arbitrary length  $L$ and finite temperature we have computed the value of $V^*$
numerically.    In Fig.~\ref{a4} we show the results. The solid black line gives the
values of $V^*$ as a function of $L$ and $T$ [the (a) and (b)  panels of Fig.~\ref{a4}
respectively]. The area below the black  curve corresponds to the range of parameters for
which the Andreev current is suppresses by the presence of a spin-splitting field, while
the area above the solid line corresponds to the range of parameters for which the
unexpected enhancement of the subgap current takes place. According to Fig.~\ref{a4}(a)
at $T=0$ the value of $V^*$ first decreases as $L$ increases, reach a minimum and then
grows again up to the asymptotic value $eV^*\approx 0.56\Delta_0$. Also the dependence of
$V^*$ on the temperature is non-monotonic having a maximum value at $T\sim 0.2 \Delta_0$.

\begin{figure}[t]
\includegraphics[width=\columnwidth]{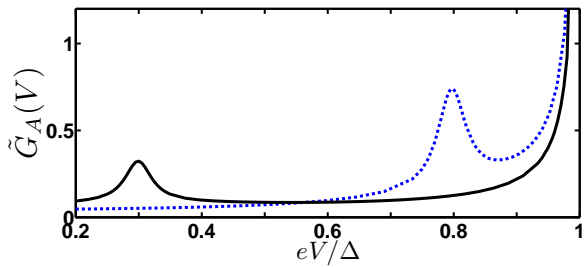}
\caption{(Color online) The bias voltage dependence of differential conductance  at $T=0$
for  fields: $h=0.3$ (black solid) and $h=0.8$ (blue dashed).   Here $\tilde{G}_A= 4 R_T
G_A $, $W=0.007$ and $L=10\xi$.} \label{a2}\vspace{-5mm}
\end{figure}

 Small spin-splitting fields, as those studied in the present work,
can be created by applying an external magnetic field $B$,  in which case $h=\mu_B B$ or
by the proximity of a ferromagnetic insulator as discussed in
Ref.~\onlinecite{Cottet2011}. It may be also an intrinsic exchange field  of weak
ferromagnetic alloys (see, for example, Ref.~\onlinecite{small_h}). Such small exchange
fields are in principle difficult to detect.  However, as we show in Fig.~\ref{a2},  by
measuring   the  subgap differential conductance $G=dI/dV$ at low temperatures, one can
accurately determine the value of $h$.  At $T=0$  the conductance shows  two well defined
peaks, one at $eV=h$ and the other at $eV=\Delta$. These are related to  a sudden increase
of the coherence length  between the  electron-hole pairs in the ferromagnet and of the
two-particle tunneling amplitude respectively. As we have seen above, at small voltages
$eV<h$ electrons with majority spins do not find time-reversed partners in the narrow
energy window around the Fermi energy, i.e. such pairs show weak coherence. By increasing
the voltage $eV \alt h$, the contribution of time-reversed electrons to the current
 gradually increases and consequently the differential conductance increases, reaching a
maximum at $eV=h$. Further increase of the voltage, $eV>h$,  leads to an  increasing
contribution to the current from non time-reversed electron-hole pairs and therefore to a
suppression of the coherent contribution to $G$.  At $h<eV\lesssim \Delta$ the
two-electron tunneling amplitude increases as $(eV-\Delta)^{-1}$ due to virtual state
contributions with energies $eV$ close to the gap; as a result the conductance shows a
sharp increase. For $h \to 0$ (normal metal) the peak moves toward $eV \to 0$ (not shown here), 
which corresponds to the zero bias peak discussed, for example, in Ref. \onlinecite{ZBA}.

In conclusion, we present an exhaustive study of the subgap charge current through
S-F-N$_{\rm e}$ hybrid structures in the presence of a spin-splitting field.  We have
demonstrated the existence of a threshold bias voltage $V^*$ above which  the Andreev
current can be enhanced by an exchange field. We also have shown that at finite
temperatures the Andreev current has a peak for values of the exchange field  close to
the superconducting gap $\Delta$. Finally, we have proposed a way to determine the
strength of small exchange fields by measuring the differential conductance. Beyond the
fundamental interest,  our results  can also be useful  for the implementation of  a
recent and  interesting proposal \cite{Cottet2011} which suggests a way to detect the  odd-triplet
component \cite{Bergeret2005} of the superconducting condensate induced in a normal metal
in contact with a superconductor and a ferromagnetic insulator. The latter  induces
an effective exchange field in the normal region. The amplitude of such induced exchange fields 
is smaller than the superconducting gap\cite{euo}. Therefore the proposed ferromagnet proximity system in Ref. \onlinecite{Cottet2011} is a candidate to observe the phenomena described in the present work.

The authors thank E.I. Kats for useful 314 discussions. F.S.B. and A.O. acknowledge the Spanish Ministry  Economy and Competitiveness under Project No. FIS2011- 316 28851-C02-02.  The work of A.O. was supported by the CSIC and the European Social Fund under the JAE-Predoc program.



\begin{thebibliography}{10}
\bibitem{footnote0} Throughout this article $\Delta$ denotes the temperature dependent BCS gap, while $\Delta_0$ its value  at $T=0$.
\bibitem{footnote1}
We neglect here the contribution to $I_A$  from partial Andreev reflection processes
at energies above the  gap. In the case of tunneling barriers, this contribution leads to
small corrections and can be neglected \cite{VBCH}.
\end{thebibliography}
\end{document}